\documentclass[10pt,conference,letterpaper]{IEEEtran}
\IEEEoverridecommandlockouts
\usepackage{cite}
\usepackage{amsmath,amssymb,amsfonts}
\usepackage{algorithmic,comment}
\usepackage{graphicx,subfigure}
\usepackage{textcomp,mathrsfs}
\usepackage{xcolor,booktabs}
\usepackage{steinmetz,relsize}
\usepackage{enumitem,balance}
\usepackage{tabularx,soul}
\usepackage{color,colortbl,textcase,bm}
\usepackage[colorinlistoftodos]{todonotes}
\usepackage{tikz}
\usetikzlibrary{positioning}
\usetikzlibrary{arrows.meta}

\def\BibTeX{{\rm B\kern-.05em{\sc i\kern-.025em b}\kern-.08emT\kern-.1667em\lower.7ex\hbox{E}\kern-.125emX}}

\definecolor{Gray}{gray}{0.92}

\usepackage{url}

\usepackage{scalerel}
\usepackage{tikz}
\usetikzlibrary{svg.path}

\definecolor{orcidlogocol}{HTML}{A6CE39}
\tikzset{
  orcidlogo/.pic={
    \fill[orcidlogocol] svg{M256,128c0,70.7-57.3,128-128,128C57.3,256,0,198.7,0,128C0,57.3,57.3,0,128,0C198.7,0,256,57.3,256,128z};
    \fill[white] svg{M86.3,186.2H70.9V79.1h15.4v48.4V186.2z}
                 svg{M108.9,79.1h41.6c39.6,0,57,28.3,57,53.6c0,27.5-21.5,53.6-56.8,53.6h-41.8V79.1z M124.3,172.4h24.5c34.9,0,42.9-26.5,42.9-39.7c0-21.5-13.7-39.7-43.7-39.7h-23.7V172.4z}
                 svg{M88.7,56.8c0,5.5-4.5,10.1-10.1,10.1c-5.6,0-10.1-4.6-10.1-10.1c0-5.6,4.5-10.1,10.1-10.1C84.2,46.7,88.7,51.3,88.7,56.8z};
  }
}

\newcommand\orcidiconA{\href{https://orcid.org/0009-0000-9566-8043}{\mbox{\scalerel*{
\begin{tikzpicture}[yscale=-1,transform shape]
\pic{orcidlogo};
\end{tikzpicture}
}{|}}}}
\newcommand\orcidiconB{\href{https://orcid.org/0000-0002-9393-7104}{\mbox{\scalerel*{
\begin{tikzpicture}[yscale=-1,transform shape]
\pic{orcidlogo};
\end{tikzpicture}
}{|}}}}

\usepackage[hidelinks]{hyperref} 

\begin{document}

\title{Transformers - Messages in Disguise\\
\thanks{\hspace{-0.4cm} $\diamond$ -- Corresponding Author.\\
\noindent\textit{\underline{NOTE:}} All vectors are denoted using lowercase, bold variables, e.g., $\mathbf{a}$, and matrices denoted using uppercase, bold variables, e.g., $\mathbf{A}$. The ``\textbf{bold}'' is removed and indices added when selecting a particular entry within a vector, e.g., $a(i)$, and matrix, e.g., $A(i,j)$.}
}


\author{\IEEEauthorblockN{Joshua {H.} Tyler$^{1\ast}$\orcidiconA, Mohamed K. M. Fadul$^{2\star}$, and Donald R. Reising$^{3\ast\diamond}$\orcidiconB}
\IEEEauthorblockA{\textit{The University of Tennessee at Chattanooga}$^{\ast}$, Chattanooga, Tennessee, USA\\
\textit{Tennessee Tech University}$^{\star}$, Cookeville Tennessee, USA\\
joshua-tyler@mocs.utc.edu$^{1}$, mfadul@tntech.edu$^{2}$, donald-reising@utc.edu$^{3}$}
}

\maketitle

\begin{abstract}
Modern cryptography, such as Rivest–Shamir–Adleman (RSA) and Secure Hash Algorithm (SHA), has been designed by humans based on our understanding of cryptographic methods. Neural Network (NN)-based cryptography is being investigated due to its ability to learn and implement random cryptographic schemes that may be harder to decipher than human-designed algorithms. NN-based cryptography may create a new cryptographic scheme that is NN-specific and that changes every time the NN is (re)trained. This is attractive since it would require an adversary to restart its process(es) to learn or break the cryptographic scheme every time the NN is (re)trained. Current challenges facing NN-based encryption include additional communication overhead due to encoding to correct bit errors, quantizing the continuous-valued output of the NN, and enabling One-Time-Pad encryption. With this in mind, the  Random Adversarial Data Obfuscation Model (RANDOM) Adversarial Neural Cryptography (ANC) network is introduced. RANDOM is comprised of three new NN layers: the (i) projection layer, (ii) inverse projection layer, and (iii) dot-product layer. This results in an ANC network that (i) is computationally efficient, (ii) ensures the encrypted message is unique to the encryption key, and (iii) does not induce any communication overhead. RANDOM only requires around $100${Kb} to store and can provide up to $2.5${Mb/s} of end-to-end encrypted communication.
\end{abstract}


\section{Introduction%
\label{s:introduction}}
\IEEEPARstart{C}{ryptography} ensures information confidentiality within modern communications systems. A system is secure if it can safeguard the confidentiality of information against all attacks.  Deep Learning (DL) is capable of addressing many challenges facing modern communications such as managing spectrum access and utilization~\cite{DARPA_SC2,Yu_ICC_2018}, aiding in the design of communication systems or designing them altogether~\cite{Shea_CR_2017,Restuccia_Mobihoc_2020,Qin_WComms_2019,Downey_Spectrum_2020}, and recognizing modulation schemes used within an operating environment~\cite{Shea_CR_2017, DARPA_RFMLS,Restuccia_DeepRadioID_2019}. Combining these capabilities with DL's ability to achieve high accuracy levels exceeding those of human-derived mathematical models even in cases where such models are not tractable or do not exist. Its ability to flourish in increasing information/data availability makes DL attractive in addressing modern communications challenges such as threats~\cite{GAO_EMOps_2021}. These capabilities also make DL attractive to adversaries who intend to use it to compromise a 
communications system's information confidentiality and integrity by persistently and consistently eavesdropping on the communications channel to build a large dataset of intercepted messages from which to extract the unencrypted message, learn the key, learn the encryption process, or combinations thereof. In Adversarial Neural Cryptography (ANC), an encryption/decryption network pair is trained with an eavesdropper attempting to monitor communications. The work in~\cite{google_arvix} introduces ANC using a Convolutional Neural Network (CNN) while highlighting the challenge of bit-recovery error without explaining why it occurs. \\
\indent This paper introduces the Random Adversarial Neural Data Obfuscation Model (RANDOM) consisting of paired encryption and decryption Neural Networks (NNs) that provide computationally efficient ANC on the edge; however, only one-directional communications are assessed for simplicity. RANDOM performs transformer-based ANC and is an improvement over current ANC approaches because it: (i) does not require quantization of the continuous-valued output of the ANC NN, (ii) incurs zero bit-recovery error, and (iii) and ensures each encrypted message is unique. Quantization must be removed because it adds communication overhead. Ensuring encrypted message uniqueness is essential for the security of the ANC network. An encrypted message must be unique to the key used to encrypt it for an ANC network to implement One-Time Pad (OTP) cryptography~\cite{rubin_otp}. Suppose encrypted messages are similar or identical across keys. In that case, it can be considered that the original encryption key is being reused, disqualifying the ANC network's use in OTP cryptography. Ensuring bit errors do not occur removes the need for error correction, such as encoding. Encoding schemes like those in IEEE Communications standards incur additional computational costs and communication overhead. RANDOM also introduces three new Neural Network (NN) layers: the projection, inverse projection, and dot-product layers. During training, these layers allow the NN to project the input data into a higher dimension for processing. Traditional NN networks reduce the data input to the network as a feature selection method; this does not work for binary data, which is already the lowest possible dimensionality. The message and key are transformed together at the higher dimension and then projected into the lower dimension, encrypted message. Lastly, RANDOM's efficiency achieves up to $2.5${Mb/s} of encrypted communication.

The rest of this paper is organized as follows. Sect.~\ref{s:related_works} summarizes previous ANC contributions and differences relative to RANDOM. Sect.~\ref{s:threat_model} introduces the threat model under which CNN, Long Short-Term Memory (LSTM), and RANDOM ANCs are trained, Sect.~\ref{s:methodology} describes RANDOM's NN layers and architecture along with the assessment experiments descriptions. Sect.~\ref{s:results} presents RANDOM results and analysis along with results and analysis corresponding to those achieved using a CNN and LSTM for comparative assessment. The paper is concluded in Sect.~\ref{s:conclusion}.
\section{Related Works%
\label{s:related_works}}
\subsection{Adversarial Neural Cryptography}
The authors of~\cite{google_arvix} introduce NN-based cryptography in the presence of an adversary known as Adversarial Neural Cryptography (ANC). ANC jointly trains an encryption and decryption network labeled Alice and Bob, respectively. An eavesdropper (a.k.a. Eve) monitors all messages encrypted by Alice to decrypt and obtain the original message. Alice is trained to encrypt the message so Bob can recover the original, unencrypted message using the encrypted message and key while ensuring Eve cannot recover the original, unencrypted message using only the intercepted, encrypted message. The authors of~\cite{google_arvix} create the ANC network using a CNN. The CNN-based ANC network does converge, but $0.5${\%} of the bits are incorrectly estimated by Bob. Alice's encryption network does not output a direct bit stream but a range of floating-point numbers ranging from $-1$ to $1$, which prevented the authors from assessing Bob's bit estimation performance under quantization. The CNN-based ANC must have the encrypted message quantized to be implemented in a communications stack. Also, the authors do not consider the computational cost of implementing their CNN-based ANC network.

The authors of~\cite{Coutinho_MDPI_2028} present a Chosen-Plaintext Attack ANC algorithm that uses an alternative training approach to influence Alice and Bob's to learn an encryption method that more closely resembles One-Time Pad (OTP) encryption. This training process begins by having Alice randomly choose one of two Eve-generated bit messages. Alice encrypts the chosen message and transmits it to Bob. Eve intercepts the encrypted message and guesses which message Alice chose. The authors of~\cite{Coutinho_MDPI_2028} propose that this training method significantly increases Alice and Bob's chance of developing an OTP-like encryption process. The authors of~\cite{Coutinho_MDPI_2028} do not state Bob's bit recovery accuracy, the computational cost of implementation, or whether the encrypted message is unique to the key used to encrypt it.
\section{Threat Model%
\label{s:threat_model}}
Similar to that of~\cite{google_arvix,Coutinho_MDPI_2028}. the communications network consists of three actors: (i) Alice, the message sender, trains RANDOM and sends RANDOM's parameters to the receiver; (ii) Bob, an authorized network user, receives Alice's encrypted messages and trained RANDOM decryption NN weights and biases; and (iii) Eve, the on-path adversary~\cite{clancy2008security}. Eve attempts to decrypt Alice's transmitted encrypted messages. Eve does this by continuously eavesdropping on the Alice-Bob communications channel and non-cooperatively, directly, and persistently detecting and intercepting Alice's transmitted, encrypted signal(s). During training, Alice models Eve using an architecture similar to Bob's, but Eve's network does not know the keys Alice used to encrypt the messages.
\section{Methodology%
\label{s:methodology}}
\subsection{Key Generation%
\label{s:key_selection}}
The encryption keys are generated from $N_{b}$ long bit sequences that (i) are bitwise symmetrical and (ii) match or fall under a set Peak Side Lobe (PSL) tolerance~\cite{Fadul_MILCOM_2021}. bitwise symmetry occurs when the number of zeros and ones are the same within a given sequence. A bit sequence's PSL is,
\begin{equation}
    p[n] = \sum_{i=0}^{N_{k}-1}k[i]\times k[i+n],
\end{equation}
where $k$ is the encryption key and $i$ circularly shifts $k$. The PSL is determined to be the largest value of $p[n]$ across all $n$.
\subsection{Data Preparation}
When training the encryption and decryption networks, all numbers that can be represented using eight bits, zero through two hundred and fifty-five, are converted into their binary representations. Eight bits allow per-byte encrypted messages to be sent without padding (i.e., adding zeros). Encryption keys are also eight bits long, with the PSL tolerance set to five. This generates seventy unique eight-bit encryption keys. During a forward pass, each encryption key is attached to each input eight-bit sequence in the minibatch to ensure all encryption keys work for all messages. Zeros are replaced with negative ones for all input eight-bit unencrypted messages and encryption keys.
\subsection{The RANDOM Architecture
\label{s:network_architecture}}
\begin{figure}[!t]
\centering
\begin{subfigure}[Projection layer process]{\label{fig:proj}\includegraphics[width=0.6\columnwidth]{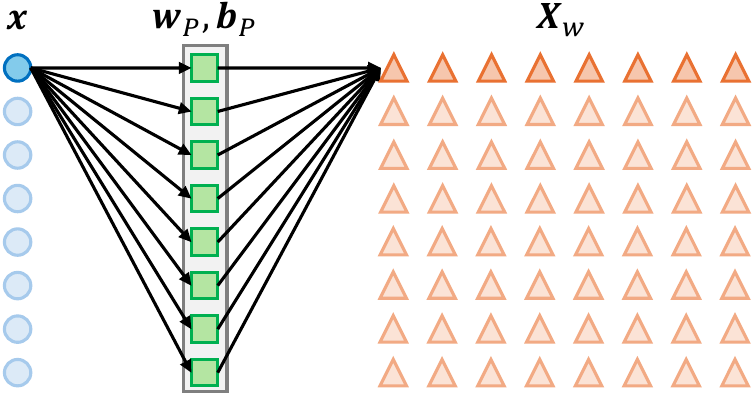}}
\end{subfigure}
\begin{subfigure}[Dot-product layer process]{\label{fig:dot_prod}\includegraphics[width=0.95\columnwidth]{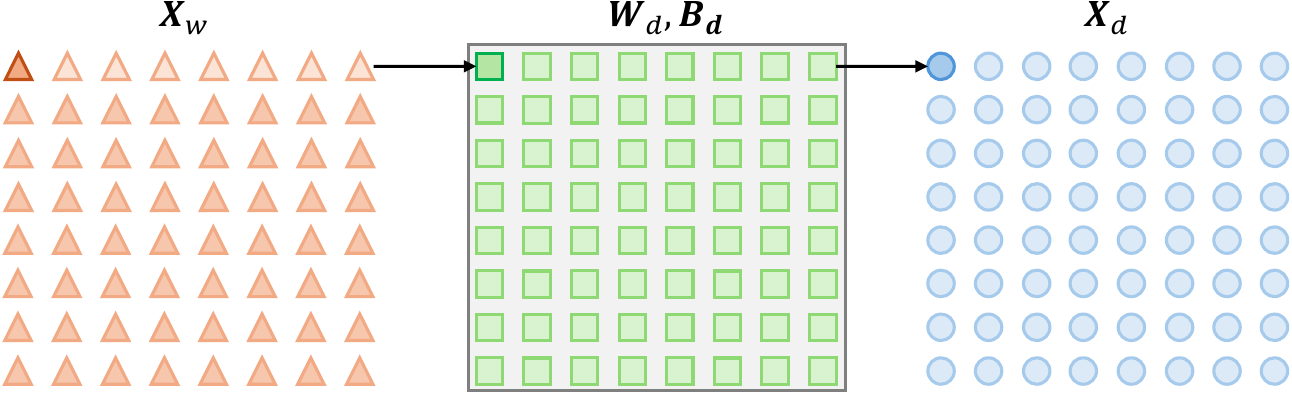}}
\end{subfigure}
\begin{subfigure}[Inverse projection layer process]{\label{fig:inv_proj}\includegraphics[width=0.6\columnwidth]{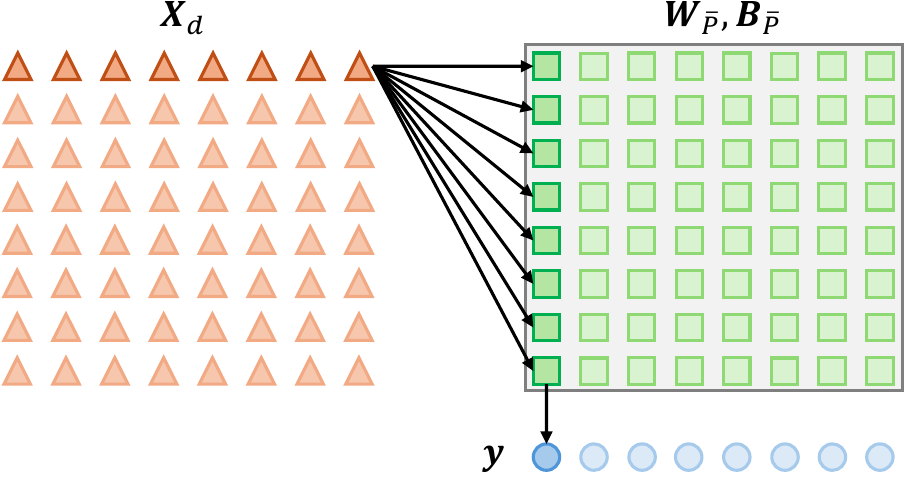}}
\end{subfigure}
\caption{Representative illustration of the projection, dot product, and inverse projection layers comprising the RANDOM. The input, $\mathbf{x}$ (blue \textcolor{blue}{$\circ$} projected into the output matrix $\mathbf{X}_{W}$(solid, orange \textcolor{orange}{$\vartriangle$}) by multiplying it with each weight (solid, green \textcolor{green}{$\square$}) within the weight matrix $\mathbf{w_{{p}}}$.}
\label{fig:network_layers}
\vspace{-5mm}
\end{figure}

The RANDOM encryption and decryption NNs are constructed of projection, dot-product, and inverse projection layers to increase the dimensionality of the input data, process the higher-dimensional data, and project the processed data back to its original dimensionality, Fig.~\ref{fig:network_layers}. The projection layer, shown in Fig.~\ref{fig:proj}, consists of weight, $\mathbf{w}_{{p}}$, and bias, $\mathbf{b}_{{p}}$ vectors. The unencrypted message vector, $\mathbf{x}$ is projected into $\mathbf{X}_{W}$ by,
\begin{equation}
    X_{W}[i,j] = x[i] \times w_{p}[j] + b_{p}[j],
\end{equation}
%
where $\mathbf{X}_{W}$ has a dimensionality of $[N_{b}{\times}N_{w}]$, $N_{b}$ is the number of unencrypted message bits (a.k.a., the length of $\mathbf{x}$), and $N_{w}$ is the projection dimension. Next, the dot-product layer, Fig.~\ref{fig:dot_prod}, is comprised of a weight and bias matrix $\mathbf{W}_{d}$ and $\mathbf{B}_{d}$. The dot-product layers operation is described by,
\begin{equation}
    \mathbf{X}_{d} = \mathbf{X}_{W} {\odot} \mathbf{W}_{{d}} + \mathbf{B}_{{d}}.
\end{equation}

\noindent The inverse projection layer, shown in Fig.~\ref{fig:inv_proj}, consists of a weight and bias matrix $\mathbf{W}_{\bar{p}}$ and $\mathbf{B}_{\bar{p}}$, respectively. $\mathbf{X}_{d}$ is projected down to $\mathbf{y}$ by,
\begin{equation}
    \mathbf{y}[i] = \sum\limits_{j=0}^{N_{w}-1}\left(\mathbf{X}_{d}[i,j]\times \mathbf{W}_{\bar{p}}[i,j] + \mathbf{B}_{\bar{p}}[i,j]\right),
\end{equation}
where $i{=}1,\dots,N_{b}$, and $N_{b}$ is the length of the encrypted message $\mathbf{y}$. All layers within each of the three networks are tanh activated. All output data is converted back from the $\pm1$ values to $[0,1]$ binary values by,
\begin{equation}
    y_{b} = \dfrac{y_{t}+1}{2},
\end{equation}
where $y$ is the tanh-activated output of a network.
\begin{figure*}[!t]
\centering
\begin{subfigure}
[Alice's encryption network]{\label{fig:alice_net}\includegraphics[width=0.32\textwidth]{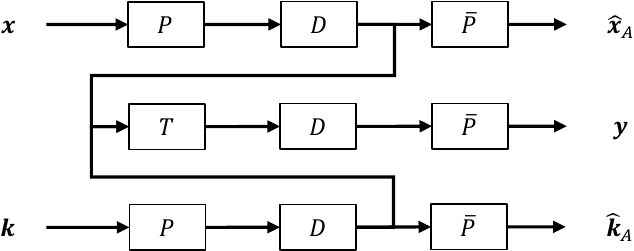}}
\end{subfigure}
\begin{subfigure}
[Bob's decryption network]{\label{fig:bob_net}\includegraphics[width=0.32\textwidth]{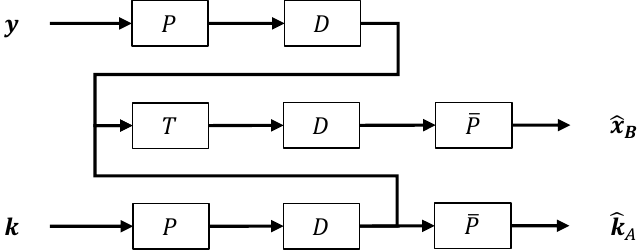}}
\end{subfigure}
\begin{subfigure}
[Eve's decryption network]{\label{fig:eve_net}\includegraphics[width=0.32\textwidth]{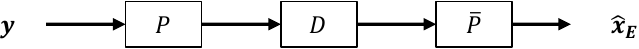}}
\end{subfigure}
\caption{RANDOM's NN architecture for Alice, Bob, and Eve in which projection, inverse projection, and dot-product layers are denoted $P$, $\bar{P}$, and $D$, respectively. A transformer operation is denoted using $T$.}
\label{fig:random_nn}
\vspace{-11pt}
\end{figure*}
\subsubsection{Encryption Network Architecture}
The encryption network takes in the unencrypted message $\mathbf{x}$, and the $8$-bit long key $\mathbf{k}$, and outputs Alice's estimation of the original message $\hat{\mathbf{x}}_{A}$, estimation of the encryption key $\hat{\mathbf{k}}_{A}$, and the encrypted output $\mathbf{y}$, as shown in Fig.~\ref{fig:alice_net}. The encryption process consists of independently projecting the unencrypted message and encryption key into a higher dimension, transforming them together, and then projecting the result back down into the encrypted message. The output of the dot-product layer is passed to the transformer layer $T$. The transformer layer computes the dot-product using the two inputs,
\begin{equation}
    \mathbf{Y}_{{T}} = \mathbf{X}_{d}\odot \mathbf{K}_{d},
\end{equation}
where $\mathbf{X}_{d}$ and $\mathbf{K}_{d}$ are the message's and key's dot-product layers outputs, respectively. The output of the transformer layer is passed to a dot-product layer and subsequent inverse projection layer. The inverse projection layer's outputs are Alice's unencrypted message estimate $\hat{\mathbf{x}}_{A}$, encrypted message $\mathbf{y}$, and key estimate $\hat{\mathbf{k}}_{A}$. The inclusion of $\hat{\mathbf{x}}_{A}$ and $\hat{\mathbf{k}}_{A}$ forces the encryption network to randomly project $\mathbf{x}$ and $\mathbf{k}$ to ensure Bob can successfully decrypt $\mathbf{y}$ using $\mathbf{k}$. 
\subsubsection{Decryption Network Architecture}
The decryption network architecture, shown in Fig.~\ref{fig:bob_net}, is similar to the encryption network's but does not have an inverse projection layer that outputs Bob's encrypted message estimate. The encrypted message $\mathbf{y}$, and key $\mathbf{k}$ input to the decryption network and its outputs are Bob's estimates of the unencrypted message $\hat{\mathbf{x}}_{B}$ and the key $\hat{\mathbf{k}}_{B}$. However, the decryption network is trained to ensure $\hat{\mathbf{k}}_{B}=\hat{\mathbf{k}}_{A}$.
\subsubsection{Adversary Network Architecture}
Eve's NN accepts only one input: the intercepted, encrypted message $\mathbf{y}_{E}$ and outputs its estimate of the original, unencrypted message sent by Alice, which is designated $\hat{\mathbf{x}}_{E}$ to differentiate it from the estimates generated by Alice and Bob. Eve's estimate of $\hat{\mathbf{x}}_{E}$ is achieved by passing $\mathbf{y}_{E}$ through a projection layer, a dot-product layer, and finally, an inverse projection layer.
\subsection{The RANDOM Loss Function}
All networks are updated during training using the Mean Squared Error (MSE) between the inputs and outputs, which is calculated as,
\begin{equation}
    \lambda(\mathbf{X},\mathbf{Y}) = \sqrt{\dfrac{1}{N_{p}\times N_{s}}\sum\limits_{p=0}^{N_{p}-1}\sum\limits_{n=0}^{N_{s}-1}\left|\mathbf{X}[p,n]-\mathbf{Y}[p,n]\right|^{2}}
\end{equation}
where $\mathbf{X}$ and $\mathbf{Y}$ are a network's input and output, $N_{p}$ is the number of predictors (a.k.a., the mini-batch size), and $N_{s}$ is a predictor's number of samples (a.k.a., the message length).
\subsection{RANDOM Training}
\subsubsection{Encryption Network Training (a.k.a., Alice)}
Alice aims to maximize the accuracy at which Bob recovers $\mathbf{x}$ using ${\mathbf{y}}$ and $\mathbf{k}$ while simultaneously preventing or inhibiting Eve's ability to reconstruct $\mathbf{x}$ using only ${\mathbf{y}}$. Alice achieves this by minimizing its overall loss, 
\begin{align}
\lambda_{A} &= \lambda_{A,A} + \lambda_{A,B} + \lambda_{A,E} \nonumber \\ 
&= [\lambda(\mathbf{x},\hat{\mathbf{x}}_{A}) + \lambda(\mathbf{k},\hat{\mathbf{k}}_{A}) + \lambda(\mathbf{y},\lfloor{\mathbf{y}}\rfloor)] + \lambda(\mathbf{x},\hat{\mathbf{x}}_{B}) \nonumber \\
&\qquad{}+ [2 - \lambda(\mathbf{x},\hat{\mathbf{x}}_{E})]
\label{eq:alice_loss}
\end{align}
where $\lambda_{A,A}$ is Alice's loss, $\lambda_{A,B}$ is Bob's loss calculated by Alice, $\lambda_{A,E}$ is Eve's loss calculated by Alice, and $\lfloor{\cdot}\rfloor$ denotes rounding of the enclosed term. Alice's calculation of $\lambda_{A,A}$, $\lambda_{A,B}$, and $\lambda_{A,E}$ ensures the encrypted message is unique for the selected key and chosen, unencrypted message while ensuring Bob and Eve can and cannot recover $\mathbf{x}$, respectively. Alice does not round $\mathbf{y}$ before transmitting it to Bob when the RANDOM AMC network is being trained. However, during RANDOM AMC network testing, Alice does round $\mathbf{y}$. Training Alice to output the encrypted message as close to $[0,1]$ values influences Alice to output a binary sequence, thus eliminating the need for quantization.
\subsubsection{Decryption Network Training (a.k.a., Bob)}
Bob is trained to recover the unencrypted message $\mathbf{x}$ from $\mathbf{y}$ using $\mathbf{k}$ and the success at which Bob can do this is determined by calculating the loss $\lambda_{B}$,  
\begin{equation}
    \lambda_{B} = \lambda(\mathbf{x},\hat{\mathbf{x}}_{B}) + \lambda(\mathbf{k},\hat{\mathbf{k}}_{B}).
\end{equation}
\subsubsection{Adversary Network Training (a.k.a., Eve)}
During the training of Alice and Bob, it is assumed that Eve has access to the unencrypted message $\mathbf{x}$, but not the key used to encrypt it. Therefore, Eve's loss function measures its network's ability to decrypt $\mathbf{y}$,
\begin{equation}
\lambda_{A} = \lambda(\mathbf{x},\hat{\mathbf{x}}_{E}).
\end{equation}
\subsection{CNN and LSTM Training}
A comparative assessment is conducted between the RANDOM, a CNN-based ANC network, and an LSTM-based ANC network. The CNN-based ANC network is implemented using the same architecture presented by the authors of~\cite{google_arvix}. The LSTM ANC network is implemented by a new architecture based on the CNN network. For Alice, the input key is appended to the end of the input message and fed into a Fully Connected Layer (FCL) with a size of $2N_{b}N_{k}$. The FCL is activated with a sigmoid layer and then fed into an LSTM layer with a $N_{b}$ size. The output of the LSTM layer is tanh activated and then passed as the encrypted message. Bob's architecture is identical to Alice's. Eve's LSTM network passes the encrypted message to an FCL with size $N_{b}$. The output of the FCL is sigmoid activated and then passed to an LSTM layer with size $N_{b}$. The output of the LSTM layer is $\tanh$ normalized and then passed as Eve's estimation of the original message. CNN and LSTM networks are trained using the same loss parameters defined in~\cite{google_arvix}. Quantization is performed by converting the $[-1,1]$-valued output of the encryption network to a $[0,2^{N_{q}}-1]$ scale where $N_{q}$ is the number of quantized steps. This is quantization is performed by,
\begin{equation}
    \mathbf{y}_{q} = \lfloor(2^{N_{q}}-1)\times (2\mathbf{y} - 1)\rfloor,
\end{equation}
where $\mathbf{y}$ is the encrypted output of Alice. These integer values are then converted to their bitwise representation. For example, a quantized value of $\mathbf{y}_{q}=2$ with a quantization level of $N_{q}=4$ will result in $[0010]$. Each binary sequence is then appended together based on their order within the quantized sequence $y_{q}$ and then transmitted to Bob. Bob returns the original sequence by converting the received bit-sequence into the integer value associated with each $N_{q}$-length bit word. Bob then converts each integer back to a $[-1,1]$ scale by,
\begin{equation}
    \mathbf{y}_{b} = \left(\dfrac{2\times\mathbf{y}_{q}}{2^{N_{q}}-1}\right)-1.
\end{equation}
Quantization is only performed when testing the CNN and LSTM network; no quantization is performed when training.
\subsection{Training Parameters: RANDOM, CNN, \& LSTM}
\subsubsection{General Training Parameters}
During training, the RANDOM, CNN, and LSTM ANC networks are updated using ADAptive Momentum (ADAM) optimization using a learning rate of $0.001$. Not all initialized networks can converge, so each network is trained for a maximum of $256$ epochs. If a network fails to converge by the $256^{\text{th}}$ epoch, that network's weights are reinitialized, and the training process restarted. All training is performed using MATLAB R2024a. The projection, dot-product, and inverse projection layers are custom NN layers. All ANC network training is performed on a high-performance four Dell Power Ridge node cluster. Each node contains two 20C/40T Intel Central Processing Units (CPUs), $192${Gb} of Random Access Memory (RAM), and four Nvidia Tesla V100 Server PCI eXpress Module (SXM) accelerators with $32${Gb} of High Bandwidth Memory 2 (HBM2) RAM. Each V100 can train ten concurrent ANC networks at a time.
\subsubsection{RANDOM ANC Specific Training Parameters}
RANDOM encryption and decryption NNs are created with internal projection dimensions of $N_{w}=[4,8,16,32]$. For each projection dimension, ten thousand training realizations are performed. A training realization consists of randomly initializing a RANDOM and training it until either the maximum number of epochs is reached or a bit recovery accuracy of $100${\%} is achieved. The NNs that converge and achieve $100${\%} bit recovery accuracy are counted. Additionally, it was observed that not all converged RANDOM NNs encrypt ${\mathbf{x}}$ or pass the unencrypted message $\hat{\mathbf{x}}$. So, Alice checks for this error by checking to see if ${\mathbf{y}}$ matches ${\mathbf{x}}$. If they match, that RANDOM is discarded, and a new RANDOM is initialized and trained.
\subsection{Inference and Throughput}
RANDOM's inference cost is measured using throughput, which is the time required to encrypt and decrypt unencrypted messages of $64$, $128$, $256$, $512$, and $1024$\footnote{The maximum message length for an IEEE Ethernet Type II frame is 1500~{bytes}; however, powers of two are used to more clearly illustrate the impact of increasing message length on throughput~\cite{802.3_Ethernet}.}~{bytes}. Throughput is assessed using a mobile and desktop chip. The chips run on the x86, Compute Unified Device Architecture (CUDA), or the Reduced Instruction Set Computer (RICS) instruction sets. Throughput is calculated as,
\begin{equation}
    \tau = N_{p}\times\dfrac{1}{t_{A}+t_{B}},
\end{equation}
where $N_{p}$ is the number of bytes comprising an unencrypted message $\mathbf{x}$, $t_{A}$ time required for Alice to encrypt the unencrypted message, and $t_{B}$ is the time needed for Bob to recover $\mathbf{x}$ using the key $\mathbf{k}$ and encrypted message $\mathbf{y}$. The x86 CPU is represented by a desktop Intel i7 13700k and a mobile Intel i9 13980HX. The CUDA chips are represented using a desktop RTX 4070Ti Super and a mobile RTX 4070. The RISC CPUs are represented using Apple's M1 Pro and an Apple A17 Pro running on an iPhone 15 Pro. All encrypting and decrypting are performed using MATLAB's forward function on \verb+dlnetwork+ classes, not \verb+seriesnetwork+ classes. All internal network functions were performed at $32${FP} on MATLAB R2024a, except for the Apple A17 pro, which ran MATLAB's iOS application for iPhone 15 Pro.
\section{Results%
\label{s:results}}
\subsection{Results: ANC Network OTP Verification}
\begin{table}[!b]
\centering
\vspace{-11pt}
\caption{Message, key, and encrypted message hexadecimal value generated by the CNN~\cite{google_arvix}, LSTM, and RANDOM-based ANC models. NOTE: Hexadecimal over bits is used to improve readability and visual conciseness.}
\includegraphics[width=\columnwidth]{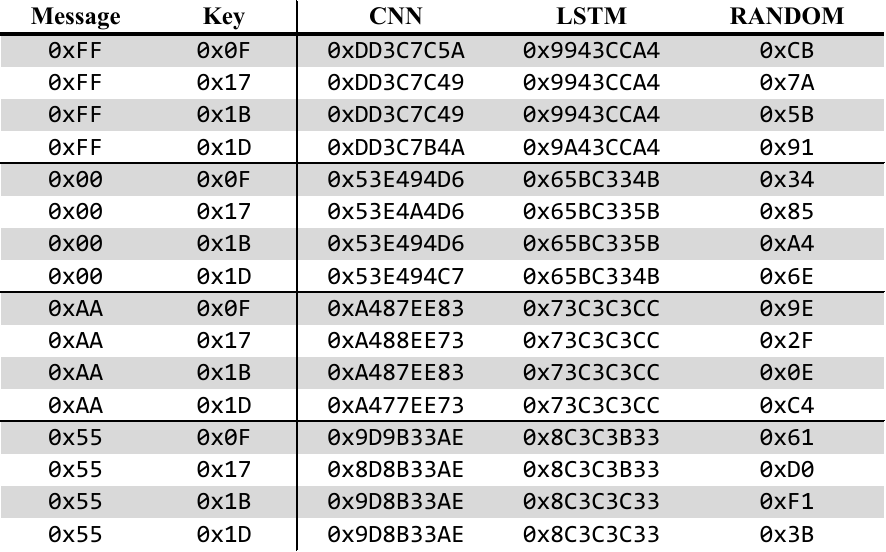}
\label{tab:message_uniqueness}
\end{table}
The CNN~\cite{google_arvix}, LSTM, and RANDOM ANC networks' ability to generate unique encrypted messages when there is a change in the unencrypted message $\mathbf{x}$, the key $\mathbf{k}$, or both is evaluated. Results are generated using four unencrypted, 8-bit messages: (i) all-ones (a.k.a., 0xFF in hexadecimal), (ii) all-zeros (a.k.a., 0x00 in hexadecimal), (ii) alternating zeros and ones where $\mathbf{x}$'s first entry is zero (a.k.a., 0xAA in hexadecimal), and (iv) alternating ones and zeros where $\mathbf{x}$'s first entry is one (a.k.a., 0x55 in hexadecimal). Each message is encrypted using four randomly selected keys from the seventy available keys compliant with the requirements in Sect.~\ref{s:key_selection}. The four chosen keys are 0x0F, 0x17, 0x1B, and 0x1D. The CNN and LSTM-based ANC networks are implemented using a quantization level of four. Using hexadecimal numbers, Table~\ref{tab:message_uniqueness} lists $\mathbf{x}$, $\mathbf{k}$, and the encrypted messages $\mathbf{y}$ generated by the three ANC networks, four unencrypted message formats, and four chosen keys.

When encrypting the all-ones message 0xFF, the CNN ANC network generates the same encrypted message using either key 0x17 or 0x1B. The LSTM ANC network generates the same encrypted message regardless of the key.

When encrypting the all-zeros message 0x00, the CNN ANC network generates the same encrypted message using either key 0x0F or 0x1B. The encrypted message is similar to those generated using keys 0x0F and 0x1B when the CNN ANC network uses key 0x17. The same encrypted message is generated when the LSTM ANC network uses either key 0x0F or 0x1D. The same encrypted message is generated whether key 0x17 or 0x1B is used. When encrypting message 0xAA, all CNN ANC network encrypted messages are similar, while the LSTM ANC network's encrypted messages are identical. When encrypting message 0x55, the CNN and LSTM ANC networks generate the same encrypted messages regardless of the key used. Compared to the CNN and LSTM ANC networks, RANDOM generates encrypted messages that change when there is a change in $\mathbf{x}$, or $\mathbf{k}$, thus ensuring OTP encryption requirements are met.

The results in Table~\ref{tab:message_uniqueness} provide an initial investigation into encrypted message uniqueness as generated by each of the ANC networks. The inquiry into encrypted message uniqueness is expanded by measuring each ANC network's average uniqueness using 17,920 encrypted messages. Each ANC network generates encrypted messages using $256$, 8-bit, and unique unencrypted messages and each of the seventy $8$-bit keys. Here ``unique'' means that no two unencrypted messages are the same. For a chosen $\mathbf{x}$, the average uniqueness of its seventy encrypted messages (i.e., one per key) is calculated as,
\begin{equation}
    u_{\mathbf{x}} = 100\times\left(\dfrac{100-s_{\mathbf{x}}}{50}\right),
\end{equation}
where,
\begin{equation}
    s_{\mathbf{x}} = \dfrac{1}{N_{b}\times N_{k}}\sum\limits_{i=0}^{N_{b}}\sum\limits_{j=0}^{N_{k}}\delta\{y[i,j]=\operatorname{mode}(y[1:N_{k},i])\},
\end{equation}
$N_{k}$ is the number of keys,$\operatorname{mode}(\cdot)$ denotes the most commonly occurring bit across the $i^{\text{th}}$ entry of the encrypted messages generated from $\mathbf{x}$ and 
all seventy keys, and $\delta\{\cdot\}$ is a logical function that returns a $1$ when the enclosed term is true and $0$ otherwise. Average encrypted message similarity $s_{\mathbf{x}}$ ranges from $50$ to $100$ where $50$ indicates \textit{no similarity} (i.e., the encrypted message is unique to the chosen key) among all $\mathbf{y}$ generated from $\mathbf{x}$ and each key. If $s_{\mathbf{x}}=100$ indicates the ANC network is outputting the same $\mathbf{y}$ regardless of the $\mathbf{k}$ used to encrypt $\mathbf{x}$. Based on $s_{\mathbf{x}}$, $u_{\mathbf{x}}$ equals $100$ when the encrypted messages are unique and $0$ when all encrypted messages are identical for selected $\mathbf{x}$ and all keys.

\begin{figure}[!t]
\centering
\includegraphics[width=0.9\columnwidth]{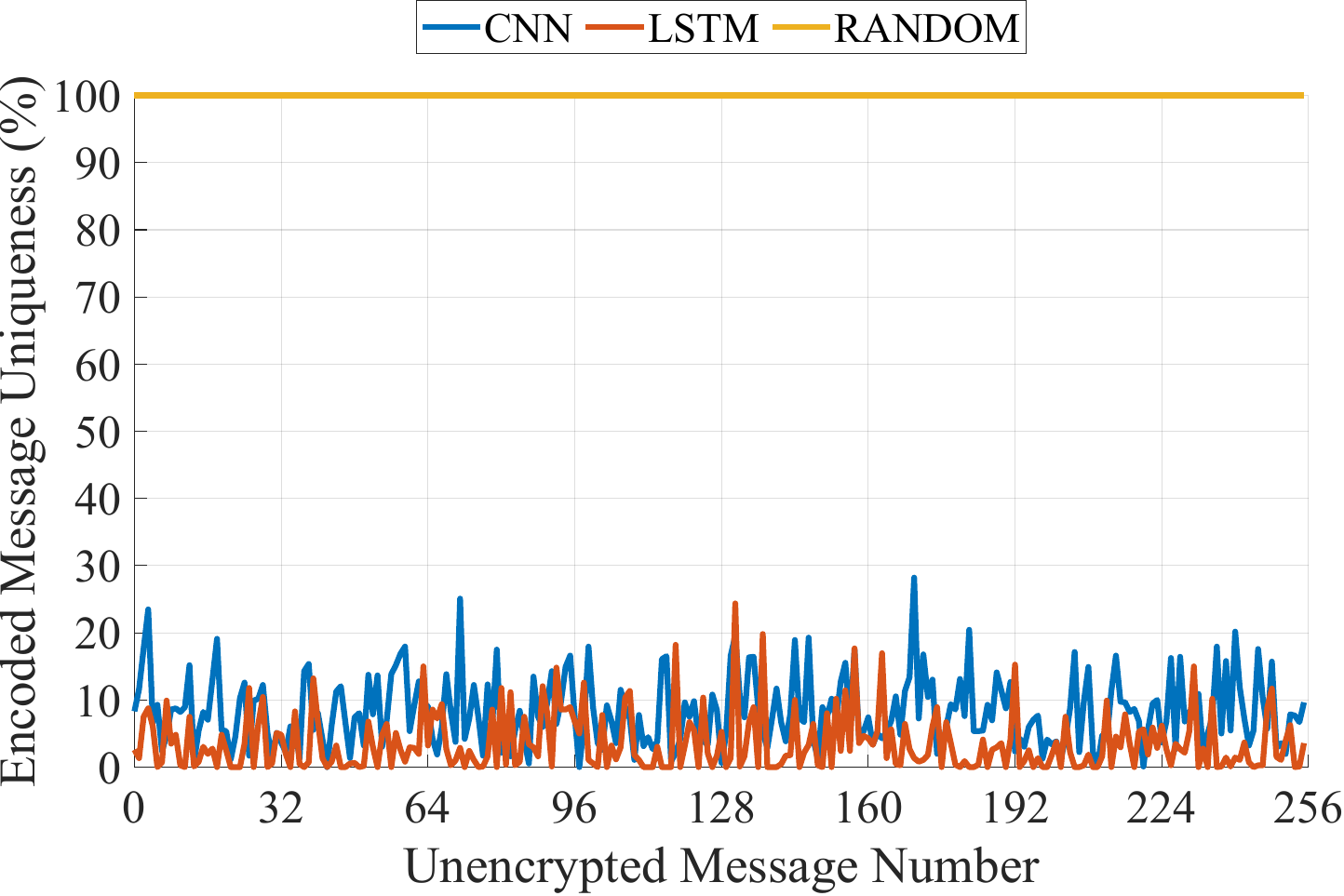}
\caption{Average encrypted uniqueness across keys for the RANDOM, CNN~\cite{google_arvix}, and LSTM ANC networks.}
\label{fig:uniqueness_plot}
\vspace{-11pt}
\end{figure}
Fig.~\ref{fig:uniqueness_plot} shows $u_{\mathbf{x}}$ for all unencrypted messages and ANC networks. RANDOM achieves a measured average uniqueness of $100${\%} for all $\mathbf{x}$ and $\mathbf{k}$. The CNN and LSTM ANC networks achieve a maximum average uniqueness above $25{\%}$ for $3$ and $1$ out of $256$ unencrypted messages, respectively. The CNN ANC network achieves an average uniqueness of $8.21${\%} across its 17,920 encrypted messages. The LSTM ANC network achieves an average uniqueness of $3.52${\%} across its 17,920 encrypted messages. The average uniqueness results show that RANDOM uses the key to fully influence encrypted message generation, ensuring that each encrypted message is unique to facilitate OTP cryptography.
\subsection{Results: Decryption Accuracy}
\begin{figure}[!b]
\centering
\vspace{-11pt}
\includegraphics[width=0.73\linewidth]{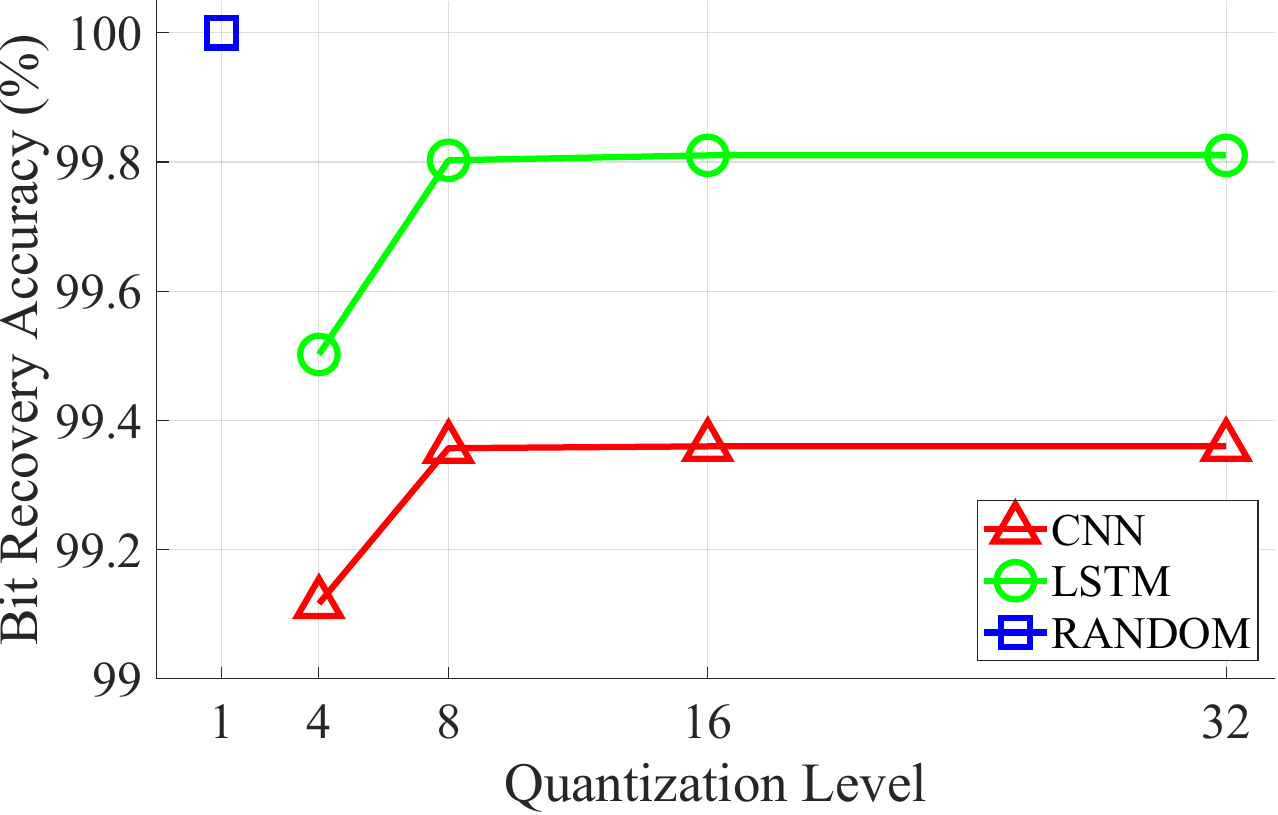}
\caption{Bit recovery accuracy when using the RANDOM  with no quantization, and the CNN~\cite{google_arvix} and LSTM network with a quantization level four, eight, sixteen, and thirty-two.}
\label{fig:bit_accuracy}
\end{figure}
Fig.~\ref{fig:bit_accuracy} shows the bit recovery accuracy between Alice and Bob when using RANDOM \textit{without} quantization and the CNN and LSTM model with quantization levels of four, eight, sixteen, and thirty-two. The results show RANDOM's ability to recover all $\mathbf{x}$ (i.e., $100${\%} accuracy) across for all $\mathbf{y}$ and $\mathbf{k}$. For a quantization level of four, the CNN and LSTM ANC networks recover $\mathbf{x}$ with an accuracy of $99.5${\%} and $99.1${\%}, respectively. For a quantization level of thirty-two, the CNN and LSTM ANC networks recover $\mathbf{x}$ with an accuracy of $99.8${\%} and $99.4${\%}, respectively. The CNN and LSTM ANC networks' recovery of $\mathbf{x}$ under quantization highlights RANDOM's advantage, which incurs no communication overhead or bit errors.
\begin{table}[!t]
\centering
\vspace{-11pt}
\caption{Average required training epochs and time to train a RANDOM encrypt and decrypt NNs with a projection dimension of $N_{w}=[4,8,16,32]$.}
\includegraphics[width=\columnwidth]{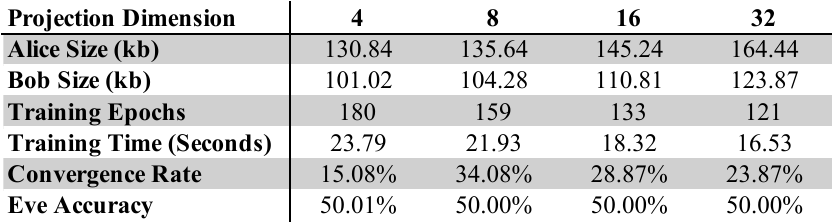}
\vspace{-11pt}
\label{tab:increasing_projection_size}
\end{table}
\subsection{Results: Projection Dimension Impacts}
\begin{figure*}
\centering
\begin{subfigure}[\textit{Best} Training Accuracy]{\label{fig:train_acc_best}\includegraphics[width=0.48\textwidth]{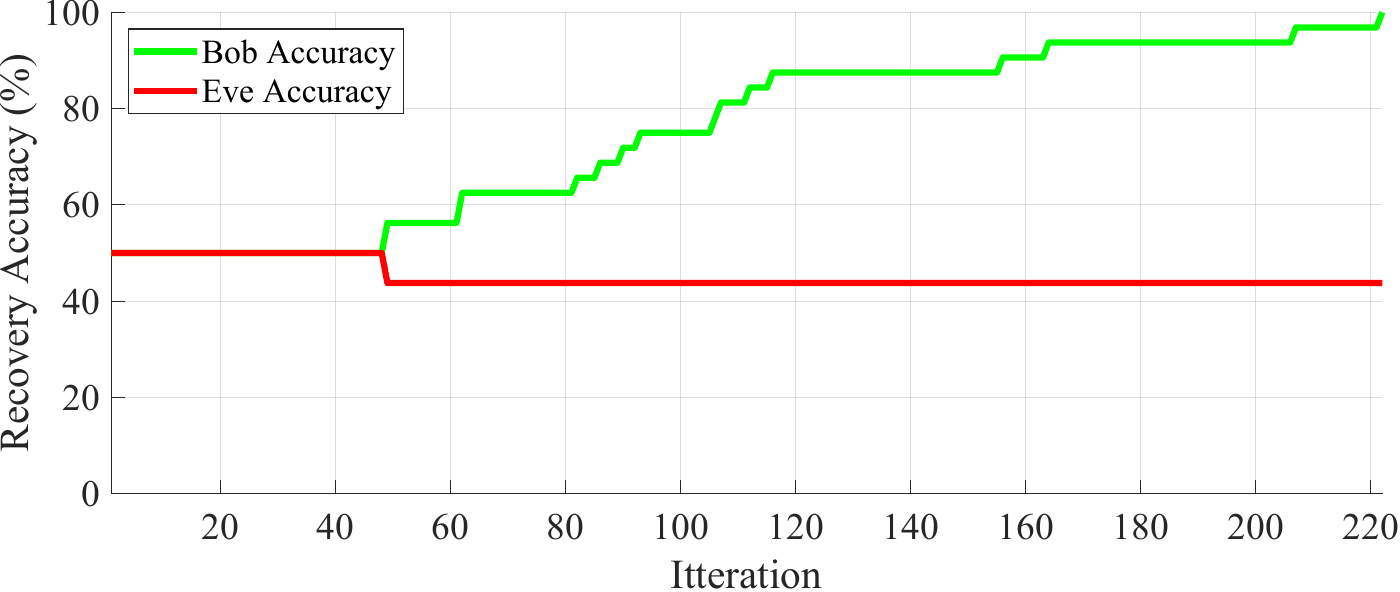}}
\end{subfigure}
\begin{subfigure}[\textit{Best} Training Loss]{\label{fig:train_loss_best}\includegraphics[width=0.48\textwidth]{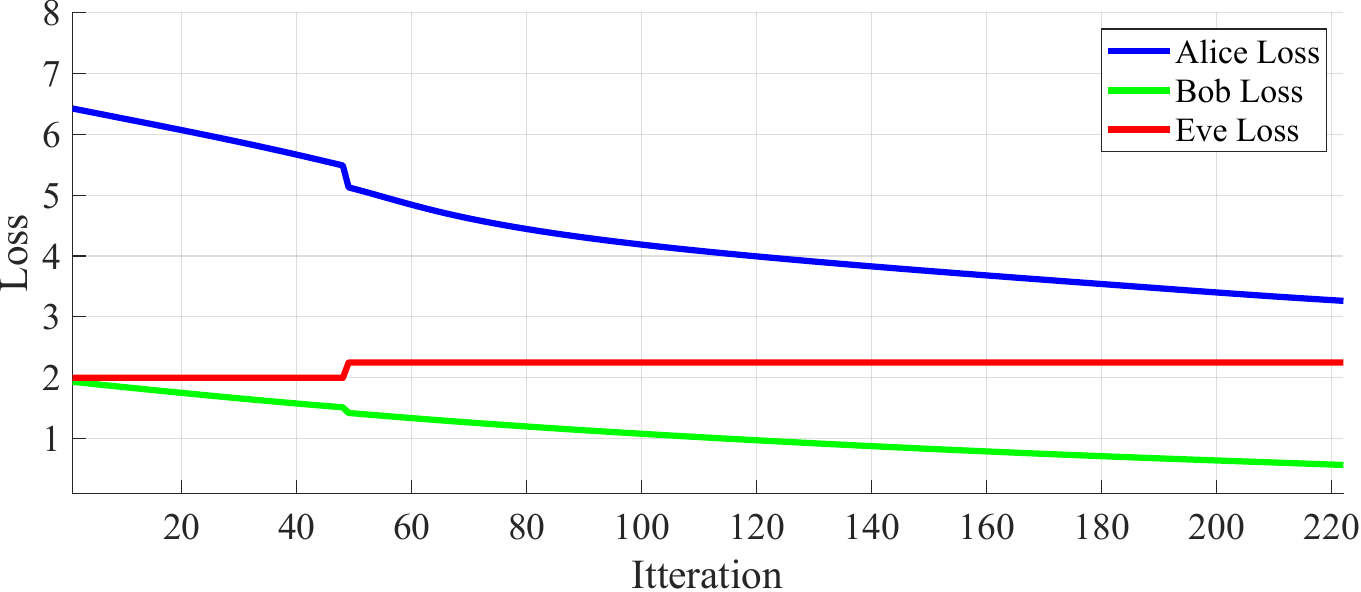}}
\end{subfigure}
\begin{subfigure}[\textit{Worst} Training Accuracy]{\label{fig:train_acc_worst}\includegraphics[width=0.48\textwidth]{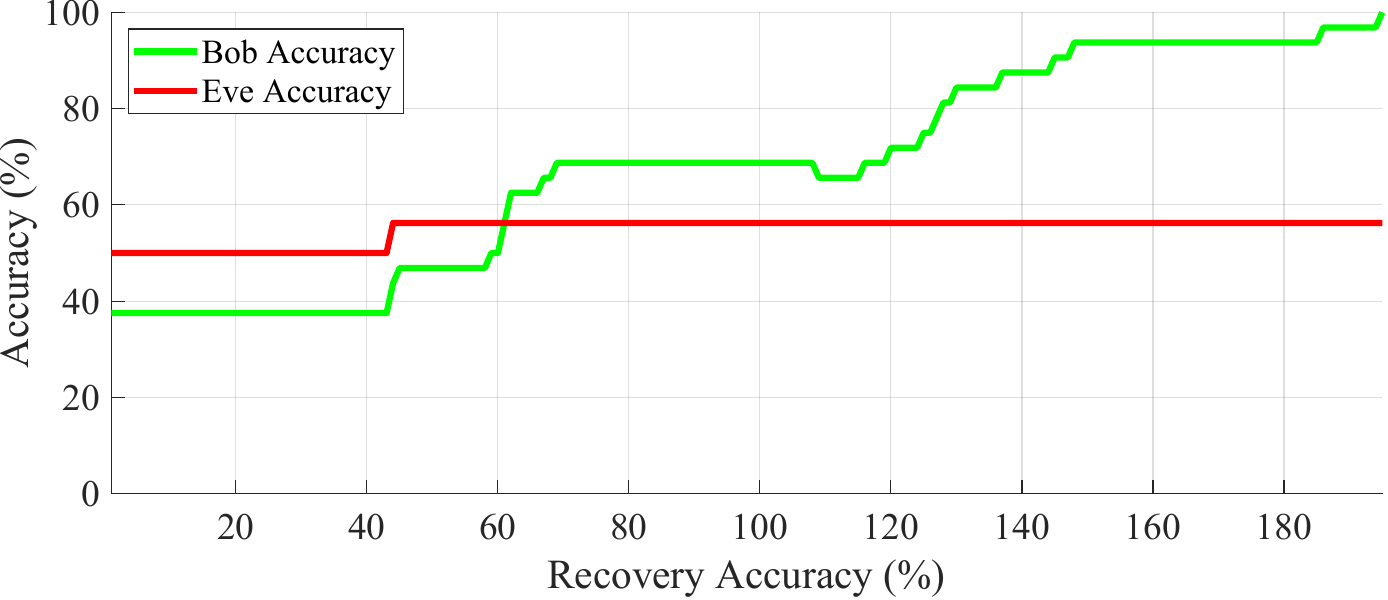}}
\end{subfigure}
\begin{subfigure}[\textit{Worst} Training Loss]{\label{fig:train_loss_worst}\includegraphics[width=0.48\textwidth]{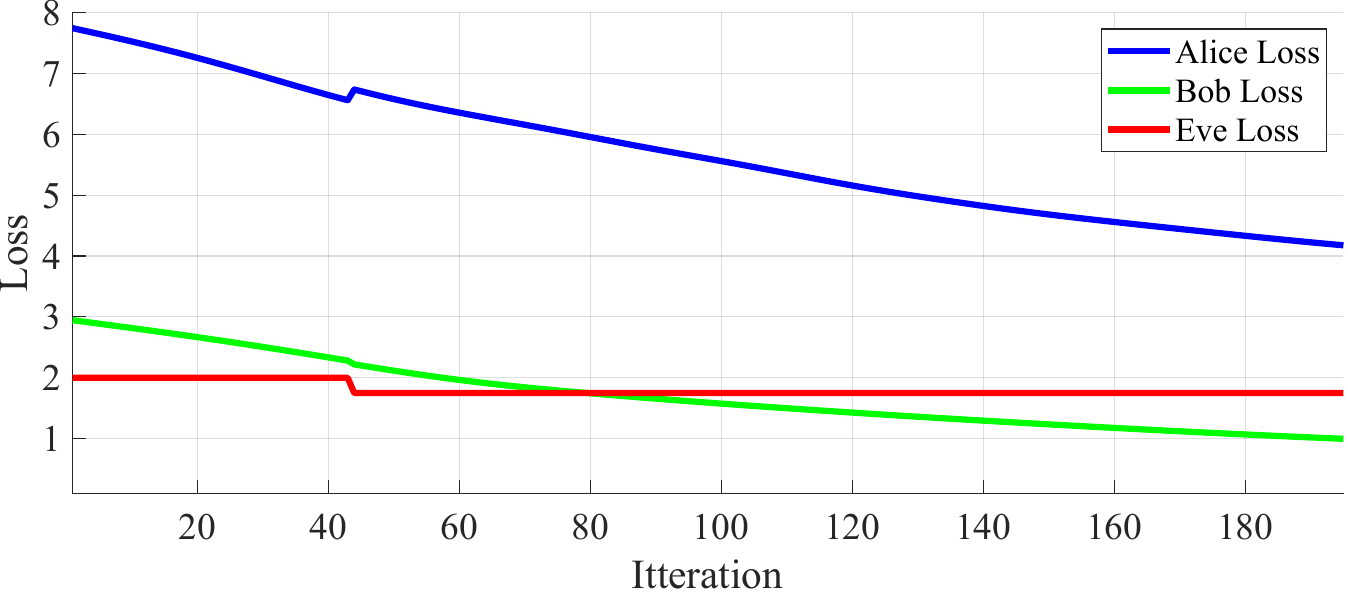}}
\end{subfigure}
\caption{Best and worst training accuracy and loss plots when training a RANDOM when all projection and inverse projection layers have a projection dimension of eight.}\label{fig:training_results}
\vspace{-5mm}
\end{figure*}
The results presented and discussed in this section focus on RANDOM's training metrics when encrypting eight-bit messages $\mathbf{x}$ using eight-bit encryption keys $\mathbf{k}$. Table~\ref{tab:increasing_projection_size} provides the training time, number of training epochs until convergence, convergence rate, RANDOM's internal projection size, and Eve's final recovery accuracy when the projection dimension $N_{w}$ equals four, eight, sixteen, or thirty-two during training.

Table~\ref{tab:increasing_projection_size} results show the number of training epochs and training time decreases as the projection dimension increases. The RANDOM networks' size does increase, but not significantly. Alice's network size rises by just over $34${kB}, going from $130${kB} to $164.44${kB}, as $N_{w}$ increases from four to thirty-two. Bob's network increases by approximately $23${kB} as $N_{w}$ increases. The convergence rate increases when $N_{w}$ increases from four to eight but drops as $N_{w}$ increases to thirty-two. The convergence rate informs the expected number of randomly initialized RANDOM ANC encrypt/decrypt pairs that must be trained before a pair converges.

Increasing the projection dimension does not affect Eve's convergence rate because it is not shown to be a factor in increasing RANDOM's ability to confuse Eve. Based on these results, the projection dimension is set to eight for the remaining experiments because it converges the fastest.

Fig.~\ref{fig:training_results} shows the ``best'' and ``worst'' training metrics for Alice, Bob, and Eve. The best training corresponds to Alice and Bob converging and recovering $\mathbf{x}$ with $100${\%} accuracy for all messages and keys, while Eve's accuracy recovering $\mathbf{x}$ is the \textit{lowest} possible (i.e., $50${\%}). The worst training corresponds to Alice and Bob converging and recovering $\mathbf{x}$ with $100${\%} accuracy for all messages and keys, while Eve's accuracy recovering $\mathbf{x}$ is the \textit{highest} possible (i.e., $100${\%}). The results in Fig.~\ref{fig:training_results} show the best and worst training accuracy and loss for Alice, Bob, and Eve.

Fig.~\ref{fig:train_acc_best} shows Bob's accuracy converges to $100${\%} by $250$ iterations. Eve's loss is $42${\%}, below the guess threshold of $50${\%}. In an ideal case, Eve's recovery accuracy would be $50${\%} rather than $0${\%}. If Eve's accuracy is $0${\%}, Eve could flip all recovered bits for $100${\%} recovery accuracy. While Eve's lowest accuracy is $42${\%}, Eve's recovery accuracy, the average across all converged RANDOM networks is $50$, shown in Table~\ref{tab:increasing_projection_size}. Fig.~\ref{fig:train_loss_best} shows Bob reaches a loss of approximately ${0.8}$ while Eve's loss is roughly $2.3$. It is worth noting that when Alice's loss value is high, Alice's loss function, Eq.~\eqref{eq:alice_loss}, accounts for Alice's, Bob's, and Eve's behavior, which means more factors contribute to the overall loss function. Fig.~\ref{fig:train_loss_best} shows Bob's loss has been reduced to $0.8$ from an initial value of $3$ while Eve's loss increases from an initial value of $2$. This shows that training Alice and Bob to be adversarial to Eve successfully prevents Eve from learning a training realization's unique encryption process.

Fig.~\ref{fig:train_acc_worst} shows the worst training realization where Alice and Bob converge while achieving a recovery accuracy of $100${\%}  bit-recovery accuracy by $197$ iterations while Eve's recovery accuracy is $58${\%} after $45$ iterations. Fig.~\ref{fig:train_loss_worst} shows the worst training realization. Interestingly, Alice's and Bob's loss is lower than that shown in Fig.~\ref{fig:train_loss_best}. Alice's loss decreases to approximately $2$ from an initial value of $5$. Bob's loss decreases from $2$ to approximately $0.5$. In contrast, Eve's loss decreases from $2$ to a constant value of $1.8$, thus showing RANDOM's ability to confuse Eve even for the worst training realization.
\subsection{Results: Inference and Throughput}
\begin{figure}[!b]
\centering
\vspace{-11pt}
\includegraphics[width=0.9\columnwidth]{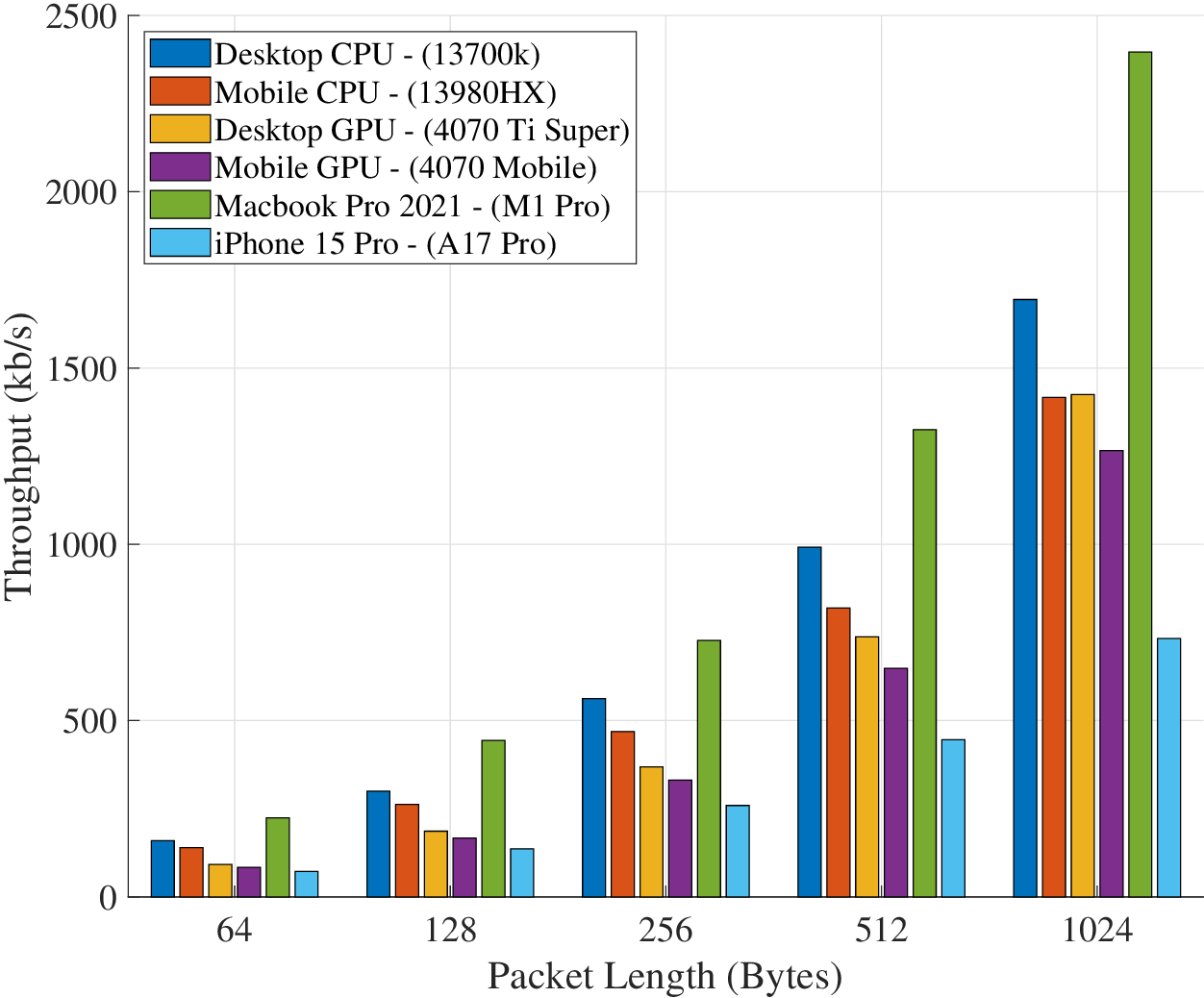}
\caption{Effective throughput across six different compute configurations when employing the RANDOM encrypt and decrypt networks.}
\label{fig:throughput}

\end{figure}
RANDOM's effective throughput is assessed using desktop and mobile implementations of three different computing architectures when encrypting/decrypting $16$, $128$, $256$, $512$, and $1024$~{byte} messages. The three computing architectures are x86, CUDA, and RISC.

Fig.~\ref{fig:throughput} shows throughput increases as the number of bytes comprising the unencrypted messages increases. This may be due to smaller message sizes not fully utilizing each computing platform's resources, whereas larger message sizes may use the platforms' mass parallelization more efficiently. Parallel processing is possible because the RANDOM performs encryption and decryption one byte at a time. When directly comparing computing platforms, Fig.~\ref{fig:throughput} shows the Apple M1 Pro achieves the highest throughput. This may be due to the $200${Gb/s} memory bandwidth of the M1 Pro being significantly higher than the Intel 13700k's $70${Gb/s} memory bandwidth. RANDOM is only $135${kb}, and while it does benefit from parallelization, it does not require the mass parallelization available on a CUDA platform. This can be seen by the desktop Intel 13700k outperforming the 4070Ti super and the mobile Intel 13980Hx outperforming the RTX 4070 mobile. In all cases except for message sizes of $1024$, the mobile 13980Hx outperforms the desktop 4070Ti Super. Therefore, RANDOM benefits more from data transfer speeds within a computing platform than raw computational capability.
\section{Conclusion%
\label{s:conclusion}}
RANDOM is introduced and compared with CNN~\cite{google_arvix} and LSTM ANC networks. RANDOM is assessed at varying projection layer dimensions, and the best and worst-case training conditions are analyzed. RANDOM is implemented on multiple consumer computing platforms to determine its effective throughput. The results show that RANDOM can guarantee that encrypted messages are unique to the key, unlike the CNN and LSTM ANC networks. The results also show that RANDOM does not require quantization, thus removing communications overhead with zero-bit errors compared to the CNN and LSTM ANC networks. When analyzing the impact of the projection dimension, it is observed that a projection dimension of eight provides the highest convergence rate. During training, Eve's best reconstruction accuracy is around $58${\%}, showing the RANDOM encryption network can consistently confuse Eve. Finally, the Apple MacBook Pro with an M1 Pro chip achieves the highest overall effective throughput. It can be reasonably concluded that due to RANDOM's lightweight nature, it benefits more from memory transfer speed versus mass parallelized computing. The highest throughput is under $2.5${Mb/s}, sufficient for encrypting real-time, encoded voice and video communications. While the data rates are adequate, they may be improved by compiling the RANDOM functionality specifically for the computing hardware used. The data rates may also be enhanced by adding specialized RANDOM inference accelerators into Network Interface Cards (NICs). Deploying RANDOM in a NIC will reduce overall data transfer between computing components, potentially improving data rates. Because RANDOM's weights only require around $100${Kb} to store, RANDOM may be hosted in an accelerator's L2 or L3 memory cache, further reducing communication overhead between components on the computing platform. Since RANDOM can provide true ANC without cross-key redundancy, future work will assess the lifespan of keys employed by a RANDOM. Future work will also investigate improving RANDOM's computational efficiency through specialized compilation and hardware accelerators.
\typeout{}

\bibliographystyle{IEEEtran}
\bibliography{002_milcom_refs}

\begin{thebibliography}{10}
\providecommand{\url}[1]{#1}
\csname url@samestyle\endcsname
\providecommand{\newblock}{\relax}
\providecommand{\bibinfo}[2]{#2}
\providecommand{\BIBentrySTDinterwordspacing}{\spaceskip=0pt\relax}
\providecommand{\BIBentryALTinterwordstretchfactor}{4}
\providecommand{\BIBentryALTinterwordspacing}{\spaceskip=\fontdimen2\font plus
\BIBentryALTinterwordstretchfactor\fontdimen3\font minus
  \fontdimen4\font\relax}
\providecommand{\BIBforeignlanguage}[2]{{%
\expandafter\ifx\csname l@#1\endcsname\relax
\typeout{** WARNING: IEEEtran.bst: No hyphenation pattern has been}%
\typeout{** loaded for the language `#1'. Using the pattern for}%
\typeout{** the default language instead.}%
\else
\language=\csname l@#1\endcsname
\fi
#2}}
\providecommand{\BIBdecl}{\relax}
\BIBdecl

\bibitem{DARPA_SC2}
{Defense Advances Research Projects Agency}, ``{Spectrum Collaboration
  Challenge --- Using AI to Unlock the True Potential of the RF Spectrum},''
  https://archive.darpa.mil/sc2/, 2017.

\bibitem{Yu_ICC_2018}
Y.~Yu, T.~Wang, and S.~C. Liew, ``Deep-reinforcement learning multiple access
  for heterogeneous wireless networks,'' vol.~37, no.~6.\hskip 1em plus 0.5em
  minus 0.4em\relax IEEE, 2019, pp. 1277--1290.

\bibitem{Shea_CR_2017}
T.~O’shea and J.~Hoydis, ``An introduction to deep learning for the physical
  layer,'' \emph{IEEE Transactions on Cognitive Communications and Networking},
  vol.~3, no.~4, pp. 563--575, 2017.

\bibitem{Restuccia_Mobihoc_2020}
F.~Restuccia and T.~Melodia, ``Polymorf: Polymorphic wireless receivers through
  physical-layer deep learning,'' in \emph{Proceedings of the Twenty-First
  International Symposium on Theory, Algorithmic Foundations, and Protocol
  Design for Mobile Networks and Mobile Computing}, 2020, pp. 271--280.

\bibitem{Qin_WComms_2019}
Z.~Qin, H.~Ye, G.~Y. Li, and B.-H.~F. Juang, ``Deep learning in physical layer
  communications,'' \emph{IEEE Wireless Communications}, vol.~26, no.~2, pp.
  93--99, 2019.

\bibitem{Downey_Spectrum_2020}
J.~Downey, B.~Hilburn, T.~O'Shea, and N.~West, ``In the future, ais—not
  humans—will design our wireless signals,'' \emph{IEEE Spectrum Magazine},
  vol. 2020, no.~5, 2020.

\bibitem{DARPA_RFMLS}
{Defense Advances Research Projects Agency}, ``{Radio Frequency Machine
  Learning Systems},''
  https://www.darpa.mil/program/radio-frequency-machine-learning-systems, 2019.

\bibitem{Restuccia_DeepRadioID_2019}
F.~Restuccia, S.~D'Oro, A.~Al-Shawabka, M.~Belgiovine, L.~Angioloni,
  S.~Ioannidis, K.~Chowdhury, and T.~Melodia, ``Deepradioid: Real-time
  channel-resilient optimization of deep learning-based radio fingerprinting
  algorithms,'' in \emph{Proceedings of the Twentieth ACM International
  Symposium on Mobile Ad Hoc Networking and Computing}, 2019, pp. 51--60.

\bibitem{GAO_EMOps_2021}
J.~W. Kirschbaum, ``{Electromagnetic Spectrum Operations: DOD Needs to Take
  Action to Help Ensure Superiority, Statement of Joseph W. Kirschbaum, PhD,
  Director, Defense Capabilities and Management, Testimony Before the
  Subcommittee on Cyber, Innovative Technologies, and Information Systems,
  Committee on Armed Services, House of Representatives},'' United States.
  Government Accountability Office, 2021.

\bibitem{google_arvix}
M.~Abadi and D.~G. Andersen, ``Learning to protect communications with
  adversarial neural cryptography,'' 2016.

\bibitem{rubin_otp}
F.~Rubin, ``One-time pad cryptography,'' \emph{Cryptologia}, vol.~20, no.~4,
  pp. 359--364, 1996.

\bibitem{Coutinho_MDPI_2028}
M.~Coutinho, R.~de~Oliveira~Albuquerque, F.~Borges, L.~J. Garcia~Villalba, and
  T.-H. Kim, ``Learning perfectly secure cryptography to protect communications
  with adversarial neural cryptography,'' \emph{Sensors}, vol.~18, no.~5, p.
  1306, 2018.

\bibitem{clancy2008security}
T.~C. Clancy and N.~Goergen, ``Security in cognitive radio networks: Threats
  and mitigation,'' in \emph{2008 3rd International Conference on Cognitive
  Radio Oriented Wireless Networks and Communications (CrownCom 2008)}.\hskip
  1em plus 0.5em minus 0.4em\relax IEEE, 2008, pp. 1--8.

\bibitem{Fadul_MILCOM_2021}
M.~K. Fadul, D.~R. Reising, K.~Arasu, and M.~R. Clark, ``Adversarial machine
  learning for enhanced spread spectrum communications,'' in \emph{MILCOM
  2021-2021 IEEE Military Communications Conference (MILCOM)}.\hskip 1em plus
  0.5em minus 0.4em\relax IEEE, 2021, pp. 783--788.

\bibitem{802.3_Ethernet}
``{IEEE Standard for Ethernet},'' \emph{{IEEE Std 802.3-2022 (Revision of IEEE
  Std 802.3-2018)}}, pp. 1--7025, 2022.

\end{thebibliography}

\end{document}